\documentclass[12pt]{article}

\def\NoBackreferences{}
\newcommand{\VersionInformation}{}  
\InputIfFileExists{Debug}{}{}

\usepackage{amsmath}
\usepackage{amsthm}
\usepackage{amsfonts}          
\usepackage{amssymb}           
\usepackage[mathscr]{eucal}
\usepackage{graphicx}
\usepackage{color}
\usepackage{hypbmsec}

\usepackage{rotating}
\usepackage{slashbox}
\usepackage[isu,bf]{caption}
\newlength{\xtrawidth}
\newlength{\xtraheight}
\usepackage[all]{xy}           

\usepackage{xr-hyper}

\ifx\NoBackreferences\EMPTYMACRO
  \usepackage[backref,linktocpage,bookmarks]{hyperref}
\else
  \usepackage[linktocpage,bookmarks]{hyperref}
\fi

\ifx\DEBUG\EMPTYMACRO
\else
  \usepackage{showlabels}
\fi

\def\clap#1{\hbox to 0pt{\hss#1\hss}}

\def\mathrlap{\mathpalette\mathrlapinternal}
\def\mathclap{\mathpalette\mathclapinternal}

\def\mathrlapinternal#1#2{%
\rlap{$\mathsurround=0pt#1{#2}$}}
\def\mathclapinternal#1#2{%
\clap{$\mathsurround=0pt#1{#2}$}}	

\makeatletter
  \def\adots{\mathinner{\mkern2mu\raise\p@\hbox{.}
      \mkern2mu\raise4\p@\hbox{.}\mkern1mu
      \raise7\p@\vbox{\kern7\p@\hbox{.}}\mkern1mu}}
\makeatother

\newcommand{\eqdef}{%
  \mathrel{\lower.1mm
    \hbox{$\stackrel{\lower.424ex\hbox{\scriptsize def}}{=}$}}
}

\newcommand{\R}{\ensuremath{{\mathbb{R}}}}
\newcommand{\C}{\ensuremath{{\mathbb{C}}}}

\newcommand{\Z}{\mathbb{Z}}
\newcommand{\CP}{\ensuremath{\mathop{\null {\mathbb{P}}}}\nolimits}

\newcommand{\iunit}{\ensuremath{\mathrm{i}}}
\newcommand{\Cunits}{\ensuremath{\C^\times}}
\newcommand{\free}{\ensuremath{\text{free}}}
\newcommand{\tors}{\ensuremath{\text{tors}}}

\DeclareMathOperator{\Span}{span}

\DeclareMathOperator{\Li}{Li}

\newcommand{\Spin}{{\mathop{\text{\textit{Spin}}}\nolimits}}

\newcommand{\textdef}[1]{{\it #1}}

\newcommand{\Xt}{{\ensuremath{\widetilde{X}}}}

\newcommand{\Ct}{{\ensuremath{\widetilde{C}}}}
\newcommand{\ZZZ}{\ensuremath{{\Z_3\times\Z_3}}}

\newcommand{\Osheaf}{\ensuremath{\mathscr{O}}}

\newcommand{\CLss}{Cartan-Leray spectral sequence}

\newcommand{\Fprepotential}{\mathscr{F}}

\newcommand{\Fprepot}[1]{\ensuremath{\Fprepotential_{{#1},0}}}
\newcommand{\FprepotNP}[1]{\ensuremath{\Fprepot{#1}^\text{np}}}
\newcommand{\FprepotX}{\Fprepot{X}}
\newcommand{\FprepotXNP}{\FprepotNP{X}}

\newcommand{\Kahler}{K\"ahler}

\newcommand{\mathemph}[1]{\textcolor{red}{\mbox{\boldmath $#1$}}}

{\end{list}}

{\everymath{\displaystyle\everymath{}}\array}%
{\endarray}

\include{pst-plot}
\psset{unit=3 pt}
\def\putlab#1)#2#3{\put#1){\makebox(0,0)[#2]{\small #3}}} 
\def\putlin#1,#2,#3,#4,#5){\put#1,#2){\line(#3,#4){#5}}}
\def\putvec#1,#2,#3,#4,#5){\put#1,#2){\vector(#3,#4){#5}}}
\def\putcx#1,#2){\put#1,#2){\circle*{1.4}}}

\newcommand{\GrantsAcknowledgements}{This research was supported in
    part by the Department of Physics and the Math/Physics Research
    Group at the University of Pennsylvania under cooperative research
    agreement DE-FG02-95ER40893 with the U.~S.~Department of Energy
    and an NSF Focused Research Grant DMS0139799 for ``The Geometry of
    Superstrings'', in part by the Austrian Research Funds FWF grant
    number P18679-N16, in part by the European Union RTN contract
    MRTN-CT-2004-005104, in part by the Italian Ministry of University
    (MIUR) under the contract PRIN 2005-023102 ``Superstringhe, brane
    e interazioni fondamentali'', and in part by the Marie Curie Grant
    MERG-2004-006374.}

{
  \makeatletter
  \newcommand{\hbs@tocstring}{}
  \newcommand{\hbs@bmstring}{}  
}

\begin{document}

\begin{titlepage}
  \vspace*{-2cm}
  \VersionInformation
  \hfill
  \parbox[c]{5cm}{
    \begin{flushright}      
      hep-th/0703134
      \\
      UPR 1179-T
      \\
      DISTA-2007
      \\
      TUW-07-05
    \end{flushright}
  }
  \vspace*{\stretch{1}}
  \begin{center}
     \Huge 
     Worldsheet Instantons, Torsion Curves\\ 
     and Non-Perturbative Superpotentials
  \end{center}
  \vspace*{\stretch{2}}
  \begin{center}
    \begin{minipage}{\textwidth}
      \begin{center}
        \large         
        Volker Braun${}^1$,
        Maximilian Kreuzer${}^2$,
        \\
        Burt A. Ovrut${}^1$, and
        Emanuel Scheidegger${}^3$
      \end{center}
    \end{minipage}
  \end{center}
  \vspace*{1mm}
  \begin{center}
    \begin{minipage}{\textwidth}
      \begin{center}
        ${}^1$ 
        Department of Physics,
        University of Pennsylvania,        
        \\
        209 S. 33rd Street, 
        Philadelphia, PA 19104--6395, USA
      \end{center}
      \begin{center}
        ${}^2$ 
        Institute for Theoretical Physics,
        Vienna University of Technology, 
        \\
        Wiedner Hauptstr. 8-10, 1040 Vienna, 
	Austria 		
      \end{center}
      \begin{center}
        ${}^3$
        Dipartimento di Scienze e Tecnologie Avanzate, 
        Universit\`a del Piemonte Orientale
        \\
        via Bellini 25/g, 15100 Alessandria, Italy, 
        and INFN - Sezione di Torino, Italy
      \end{center}
    \end{minipage}
  \end{center}
  \vspace*{\stretch{1}}
  \begin{abstract}
    \normalsize 
    As a first step towards computing instanton-generated
    superpotentials in heterotic standard model vacua, we determine
    the Gromov-Witten invariants for a Calabi-Yau threefold with
    fundamental group $\pi_1(X)=\ZZZ$. We find that the curves fall
    into homology classes in $H_2(X,\Z)=\Z^3\oplus(\Z_3\oplus\Z_3)$.
    The unexpected appearance of the finite \emph{torsion} subgroup in
    the homology group complicates our analysis.  However, we succeed
    in computing the complete genus-$0$ prepotential.  Expanding it as
    a power series, the number of instantons in any integral homology
    class can be read off.  This is the first explicit calculation of
    the Gromov-Witten invariants of homology classes with torsion. We
    find that some curve classes contain only a single instanton. This
    ensures that the contribution to the superpotential from each such
    instanton cannot cancel.
  \end{abstract}
  \vspace*{\stretch{5}}
  \begin{minipage}{\textwidth}
    \underline{\hspace{5cm}}
    \centering
    \\
    Email: 
    \texttt{vbraun}, \texttt{ovrut@physics.upenn.edu}, 
    \texttt{Maximilian.Kreuzer@tuwien.ac.at},
    \texttt{esche@mfn.unipmn.it}
  \end{minipage}
\end{titlepage}

\section{Introduction}
\label{sec:intro}

One of the challenges of string theory is to stabilize all moduli.
There has been considerable progress in this direction in recent
years, particularly for closed string moduli in type II
compactifications~\cite{Curio:2001qi, Kachru:2003aw,
  Balasubramanian:2005zx, Choi:2005ge, Cvetic:2007ju}. Comparably less
progress has been made in stabilizing the moduli of $N=1$
supersymmetric $E_{8} \times E_{8}$ heterotic string vacua.
Non-vanishing flux can remove some moduli, most notably those
associated with deformations of the complex structure. However,
stabilizing the remaining moduli is made difficult by the fact that it
is not possible to generate a \emph{perturbative} superpotential for
them. This follows from $N=1$ supersymmetry and the fact that,
classically, deformations of the Calabi-Yau threefold and holomorphic
vector bundle are unobstructed.  If there were only perturbative
corrections, this would rule out $N=1$ supersymmetric heterotic string
compactifications, even those with non-zero flux. However, the
complete string theory admits \emph{non-perturbative} corrections as
well. The best understood of these is the superpotential generated by
gaugino condensation in the strongly coupled hidden
sector~\cite{Lukas:1997rb}. Gaugino
condensation can help to fix at least some of the remaining moduli.
However, this can never stabilize them all.  In particular, the moduli
associated with the deformations of the holomorphic vector bundle do
not enter either the flux or the gaugino condensate superpotentials.
Hence, the potential energy remains flat in those directions.

To fix the remaining moduli in $N=1$ supersymmetric heterotic vacua,
one must consider the non-perturbative superpotential generated by
worldsheet instantons. It was shown in~\cite{Lima:2001nh, Lima:2001jc,
  Buchbinder:2002ic, Buchbinder:2002pr, Buchbinder:2002wz,
  Buchbinder:2003pi, Buchbinder:2002ji} that a string wrapped on a
single genus-$0$ holomorphic curve, $C$, in the background Calabi-Yau
threefold generates a non-vanishing contribution to the
superpotential. This contribution is of the form $\text{const}\cdot
P({\phi})\cdot \exp{\int_{C}\omega}$, where $P(\phi)$ is the
``Pfaffian'', typically a homogeneous polynomial of the vector bundle
moduli $\phi$, and $\omega$ is the \Kahler{} form. Note that the
vector bundle moduli enter the superpotential through these worldsheet
contributions and, at least in principle, can now be stabilized. This
was demonstrated within the context of a simplified theory
in~\cite{Buchbinder:2003pi, Buchbinder:2004im, Buchbinder:2005jy}.
Clearly, computing the complete instanton superpotential requires one
to sum over all such contributions and, hence, to count and classify
all holomorphic genus-$0$ curves. It is well-known that these are
specified by the ``instanton numbers'', or Gromov-Witten invariants,
of the Calabi-Yau threefold.

There are two approaches towards computing these invariants. For some
manifolds, one can directly compute the instanton corrections to the
prepotential of the A-model topological string theory. For Calabi-Yau
manifolds with a ``mirror'' threefold, a second method is much
easier~\cite{Candelas:1990rm}.  First, the classical prepotential of
the B-model topological theory on the mirror manifold is computed.
Mirror symmetry identifies this with the prepotential of the A-model
on the original Calabi-Yau space. One finally expands this in a power
series to find the instanton numbers.  Using these methods, the
instanton numbers of a variety of Calabi-Yau manifolds have been
computed, such as toric complete intersection threefolds.

Once the Gromov-Witten invariants of the Calabi-Yau threefold of a
heterotic vacuum are known, one can attempt to compute the complete
instanton superpotential. To do this, the holomorphic vector bundle
must be specified. For generic heterotic vacua, the calculation of the
superpotential is difficult. Hence, it has only been fully carried out
under very restrictive conditions, such as on a toric complete
intersection Calabi-Yau manifold with the vector bundle induced from
the ambient space. Rather remarkably, in all of these simple cases,
the instanton-generated superpotential vanishes~\cite{Distler:1986wm,
  Distler:1987ee, Silverstein:1995re, Basu:2003bq, Beasley:2003fx}.
This cancellation occurs in a very distinctive way. The contribution
to the superpotential of each curve in a fixed homology class contains
the same factor $\exp{\int_{C}\omega}$.  Therefore, when one sums over
the curves in this class the exponential factors out leaving a sum
over the Pfaffian components.  In the restrictive cases analyzed
in~\cite{Distler:1986wm, Distler:1987ee, Silverstein:1995re,
  Basu:2003bq, Beasley:2003fx}, this sum of Pfaffians cancels.  Hence,
the complete instanton induced superpotential vanishes.

In this paper, we begin the calculation of the non-perturbative
superpotential induced by the worldsheet instantons of a very
different class of heterotic string vacua, namely, the ``heterotic
standard model'' compactifications presented in~\cite{Braun:2005nv,
  Braun:2005ux, Braun:2005bw, Braun:2005zv, Braun:2006me,
  Braun:2006ae, Braun:2006th}. These theories have the exact matter
spectrum of the minimal supersymmetric standard model in the visible
sector, and a relatively small number of both geometric and vector
bundle moduli. The associated Calabi-Yau manifold, $X$, is the
quotient of an elliptically fibered threefold on which the finite
group $\ZZZ$ acts freely. Hence, $\pi_1(X)=\ZZZ$.  This does not fall
into the category of manifolds, such as toric complete intersection
spaces, considered in the previous paragraph. Furthermore, the
holomorphic vector bundle is constructed by extension and is not
related to the tangent bundle of the Calabi-Yau space nor is it coming
from the ambient space. It follows that the vanishing theorems do not
apply. Indeed, various arguments lead to the conclusion that the
complete instanton superpotential of heterotic standard model vacua is
non-zero.  Nevertheless, there remains a possibility of cancellations
between the different instantons in each given curve class. Therefore,
the complete superpotential in this context must be computed with
great care.

To evaluate the instanton superpotential of heterotic standard model
vacua, one must first compute the degree-2 integer homology of $X$.
We find that
\begin{equation}
  H_2\big(X,\Z\big) 
  = 
  H_2\big(X,\Z\big)_\free \oplus H_2\big(X,\Z\big)_\tors
  = 
  \Z^3 \oplus \big( \Z_3 \oplus \Z_3 \big)
  .
\end{equation}
That is, the homology classes of curves are not just a \textdef{free}
rank $3$ lattice, but contain a \textdef{torsion} subgroup
$\Z_3\oplus\Z_3$. Non-vanishing torsion subgroups in the homology of
Calabi-Yau spaces have been calculated in a small number of differing
contexts ~\cite{Aspinwall:1995rb, Batyrev:2005jc, gross-2005,
  Ferrara:1995yx}. As discussed below, it is essential when computing
the complete instanton superpotential that the torsion subgroup of the
degree-$2$ integral homology be explicitly computed.

Second, one must compute the number of holomorphic rational
curves\footnote{That is, $\CP^1 \subset X$. Due to holomorphy, higher
  genus curves do not generate a superpotential.} on the Calabi-Yau
threefold $X$. As discussed previously, this is accomplished by
calculating the genus-$0$ prepotential in the associated topological
string theories. However, for the non-simply connected, elliptic
Calabi-Yau threefolds of heterotic standard model vacua, this
calculation is complicated by two facts. First, it is usually assumed
that mirror symmetry will allow one to completely solve for the
instanton numbers. However, trying to do the computation in the
present context one quickly realizes that this is not so. Since $X$ is
not a complete intersection, one cannot simply apply the toric
algorithms. Second, no one has previously computed the topological
string prepotential for torsion curves. We have solved both of these
problems. First consider the A-model.
\begin{itemize}
\item Quotienting the prepotential on the covering space $\Xt$ given
  in~\cite{Hosono:1997hp}, we can compute much of the prepotential on
  $X=\Xt/(\ZZZ)$. This prepotential includes torsion information, but
  is only valid to linear order in one of the \Kahler{} moduli.
\item The same part of the prepotential on $X$ can be computed
  directly by counting curves on $X$, with complete agreement.
\end{itemize}
These results can be extended using the B-model and mirror
symmetry.
\begin{itemize}
\item Mirror symmetry for the toric complete intersection $\Xt$ is an
  algorithm to compute the prepotential. However, there are many
  non-toric divisors. Therefore, after descending to $X$ the
  prepotential is valid to all orders in the \Kahler{} moduli but the
  torsion information is lost.
\item We show that $\Xt$ is self-mirror. It turns out that in the
  mirror representation all moduli are toric, so that we can give a
  numerical algorithm for calculating the instanton numbers on $X$ to
  any degree in the \Kahler{} moduli and including all torsion
  information.
\end{itemize}
We will present the calculation of the integer homology, including
torsion, in~\cite{PartA}. In that paper, the prepotential and the
associated Gromov-Witten invariants will be computed purely within the
context of the A-model. The B-model calculations will be given in a
second paper~\cite{PartB}. Here, we simply summarize the arguments and
present the final results.

Having computed the instanton numbers, the final step is to introduce
the holomorphic vector bundles associated with heterotic standard
model vacua. Specifying this, one calculates the non-perturbative
superpotential induced by each holomorphic curve and then sums over
all curves in a given homology class and over all homology classes.
This will be attempted in future publications. However, we can already
state one important result. Namely, we will find that in some curve
homology classes on $X$ there is a single\footnote{On the quintic,
  quite distinctly, each effective curve class contains thousands of
  instantons.} instanton, which locally looks like
$\Osheaf_{\CP^1}(-1)\oplus \Osheaf_{\CP^1}(-1)$.  Therefore, its
contribution cannot cancel against other instanton contributions.

\section{The Calabi-Yau threefold}
\label{sec:CY}

To obtain phenomenological low energy gauge groups
within the context of the $E_8\times E_8$ heterotic string, it is
necessary to spontaneously break the visible sector $E_8$. One way to
do this, which was used in the context of heterotic standard
models~\cite{Braun:2005ux, Braun:2005bw, Braun:2005zv, Braun:2005nv},
is to first embed an $SU(4)$ gauge instanton in the $E_8$, breaking it
initially to its commutant $\Spin(10)$. One then adds a $\ZZZ$ Wilson
line, breaking the gauge group further to
\begin{equation}
  E_8 \longrightarrow \Spin(10) \longrightarrow
  SU(3)_C \times SU(2)_L \times U(1)_Y \times U(1)_{B-L}
  .
\end{equation}
In order to turn on discrete Wilson lines, the manifold must have
sufficiently many non-contractible curves. Hence, we are interested in
a Calabi-Yau threefold $X$ with $\pi_1(X)=\ZZZ$.

In this paper, we consider a slightly different quotient than the
one in~\cite{Braun:2004xv}.  This is done so as to simplify the
application of toric mirror symmetry.  However, in practice there is
little difference between the two spaces and it is a simple exercise
to extend the results of this paper to the heterotic standard model
manifold. The threefold we consider here is the free $\ZZZ$ quotient
\begin{equation}
  X = \Xt \big/ \big(\ZZZ\big)
\end{equation}
of the complete intersection
\begin{equation}
  \label{eq:eqXt}
  \begin{split}
    \Xt 
    \;&= 
    \left\{
      \begin{array}{c}
        t_0 \Big( x_0^3+x_1^3+x_2^3 \Big) + 
        t_1 \Big( x_0 x_1 x_2 \Big)
        = 0 \\
        \big( \lambda_1 t_0 + t_1\big)
        \Big( y_0^3+y_1^3+y_2^3 \Big) 
        +
        \big( \lambda_2 t_0 + \lambda_3 t_1\big)
        \Big( y_0 y_1 y_2 \Big)
        = 0
      \end{array}
    \right\}
    \\
    &\subset
    \CP^2_{[x_0:x_1:x_2]} \times 
    \CP^1_{[t_0:t_1]} \times 
    \CP^2_{[y_0:y_1:y_2]}
    .    
  \end{split}
\end{equation}
The defining equations allow for three complex
parameters\footnote{These $3$ parameters are the $h^{21}(X)=3$ complex
  structure moduli.} $\lambda_1$, $\lambda_2$, $\lambda_3$ respecting
the free $\ZZZ$ group action on $\Xt$,
\begin{equation}
  \begin{split}
    g_1:&\;
    \begin{cases}
      [x_0:x_1:x_2] \mapsto
      [x_0:\zeta x_1:\zeta^2 x_2]
      \\
      [t_0:t_1] \mapsto
      [t_0:t_1] 
      ~\text{(no action)}
      \\
      [y_0:y_1:y_2] \mapsto
      [y_0:\zeta y_1:\zeta^2 y_2]
    \end{cases}
    \\
    g_2:&\;
    \begin{cases}
      [x_0:x_1:x_2] \mapsto
      [x_1:x_2:x_0]
      \\
      [t_0:t_1] \mapsto
      [t_0:t_1] 
      ~\text{(no action)}
      \\
      [y_0:y_1:y_2] \mapsto
      [y_1:y_2:y_0]
      ,
    \end{cases}
  \end{split}
\end{equation}
where $\zeta=e^{2\pi\iunit/3}$ is a third root of unity.  The Hodge
numbers and homology groups of the covering space $\Xt$ have been
worked out previously~\cite{MR923487, Ovrut:2002jk}. They are
\begin{equation}
  \label{eq:Xthomology}
  h^{p,q}\big(\Xt\big) = ~
  \vcenter{\xymatrix@!0@=7mm@ur{
    1 &  0  &  0  & 1 \\
    0 &  19 &  19 & 0 \\
    0 &  19 &  19 & 0 \\
    1 &  0  &  0  & 1
    ,
  }}
  \qquad
  H_i\big(\Xt, \Z\big) 
  =
  \begin{cases}
    \Z
    & i=6 \\
    0
    & i=5 \\
    \Z^{19}
    & i=4 \\
    \Z^{40}
    & i=3 \\
    \Z^{19} 
    & i=2 \\
    0
    & i=1 \\
    \Z
    & i=0
    . 
  \end{cases}
\end{equation}
The homology of the quotient $X=\Xt/(\ZZZ)$ is more complicated to determine. 
We will investigate it in
the following section. But before doing so, note that there are nine
particularly simple genus-$0$ curves in $X$, which turn out to be
important. Inspecting the defining equations, we see that they are
identically satisfied if
\begin{equation}
  x_0^3+x_1^3+x_2^3 = 0 = x_0 x_1 x_2
  , \quad
  y_0^3+y_1^3+y_2^3 = 0 = y_0 y_1 y_2  
  .
\end{equation}
Since two cubics intersect in nine points, there are nine such
solutions $[x_0^{(i)}:x_1^{(i)}:x_2^{(i)}]$ for the first equation and 
nine independent solutions $[y_0^{(j)}:y_1^{(j)}:y_2^{(j)}]$ for the
second equation. This yields $9\cdot 9 = 81$ genus-$0$ curves 
\begin{equation}
  \Ct_{ij}: \CP^1 \to \Xt,~
  [t_0:t_1] \mapsto 
  \Big( 
  [x_0^{(i)}:x_1^{(i)}:x_2^{(i)}],
  [t_0:t_1], 
  [y_0^{(j)}:y_1^{(j)}:y_2^{(j)}]
  \Big)
\end{equation}
on $\Xt$, each wrapping the $\CP^1_{[t_0:t_1]}$. The $\ZZZ$ group action
identifies them in $9$-tuples and, hence, they define $9$ genus-$0$
curves
\begin{equation}
  \label{eq:si}
  \big\{s_0,\dots, s_8\big\} 
  = \Big\{ \Ct_{ij} / (\ZZZ) \Big\}
\end{equation}
on the quotient threefold $X$.

\section{Quotient Homology}
\label{sec:quot}

How can we determine the homology of the quotient $X$ in terms of the
homology of the covering space $\Xt$ and, in particular, where does
the torsion part $H_2(X,\Z)_\tors=\Z_3\oplus\Z_3$ come from? The most
complete answer to this question would be the \CLss{}. However, one
can understand $H_2(X,\Z)$ much more simply as follows.

Each curve on the covering space yields a curve on the quotient 
by taking its image under the quotient map $q:\Xt\to X$.
Now for each rational curve $\Ct$ on $\Xt$, there are $8$
other\footnote{There is no analytic free map $\CP^1\to\CP^1$ without
  fixed points, so the free group action must map genus-$0$ curves to
  different genus-$0$ curves.} images $g \Ct$, $g\in \ZZZ$ under the
group action. Clearly, their projection to the quotient is the same,
\begin{equation}
  q\big( \Ct \big) = q\big( g \Ct \big)
  .
\end{equation}
Hence, one should impose the relations $\Ct-g \Ct=0$ on the homology
of $\Xt$ to learn about the homology of $X$. This is called the
\textdef{coinvariant homology} of $\Xt$. We find that
\begin{equation}
  H_2\big(\Xt,\Z\big)_{\ZZZ} = 
  H_2\big(\Xt,\Z\big) \Big/ 
  \Span\big\{ \Ct - g \Ct \big\} = 
  \Z^3 \oplus \Z_3 \oplus \Z_3
  . 
\end{equation}
It turns out that these relations contain $3(s_i-s_j)$, but do not
include $s_i-s_j$. This is the origin of torsion in the quotient.
The $8$ non-trivial torsion homology classes can be represented by the
differences $s_i-s_0$, $i=1,\dots,8$.

In general, the coinvariant homology is only one ingredient in the
\CLss{} and need not coincide with the homology of the quotient.
However, in our case this turns out to be sufficient for the curve
classes~\cite{PartA}. Hence, we obtain
\begin{equation}
  H_2\big(X,\Z\big) = 
  H_2\big(\Xt,\Z\big)_{\ZZZ} = 
  \Z^3 \oplus \Z_3 \oplus \Z_3 .
\end{equation}
In summary, the Hodge numbers and homology groups of the quotient
$X=\Xt/\ZZZ$ are
\begin{equation}
  \label{eq:Xhomology}
  h^{p,q}\big(X\big) = ~
  \vcenter{\xymatrix@!0@=7mm@ur{
    1 &  0  &  0  & 1 \\
    0 &  3  &  3  & 0 \\
    0 &  3  &  3  & 0 \\
    1 &  0  &  0  & 1
    ,
  }}
  \qquad
  H_i\big(X, \Z\big) 
  =
  \begin{cases}
    \Z
    & i=6 \\
    0
    & i=5 \\
    \Z^3 \oplus \Z_3\oplus\Z_3
    & i=4 \\
    \Z^8 \oplus \Z_3\oplus\Z_3
    & i=3 \\
    \Z^3 \oplus \Z_3\oplus\Z_3 
    & i=2 \\
    \Z_3\oplus\Z_3
    & i=1 \\
    \Z
    & i=0
    .
  \end{cases}
\end{equation}
Note that these results are in accord with the self-mirror
property~\cite{PartB} and the conjecture of~\cite{Batyrev:2005jc} that
the torsion parts $H_1(X,\Z)_\tors$ and $H_2(X,\Z)_\tors$ are
exchanged under mirror symmetry.

\section{The Prepotential}
\label{sec:prepotb}

Having found the homology classes of curves, we are ready to
count the number of genus-$0$ curves in each homology class. This
is achieved by calculating the worldsheet instanton corrections to the
genus-$0$ prepotential in the A-model topological string on $X$. Formally, the
non-perturbative prepotential is given by
\begin{equation}
  \FprepotX^{np} =  
  \sum_{C\in H_2(X,\Z)} n_C \; 
  \Li_3\Big( e^{2 \pi \iunit \int_C \omega} \Big),
  \label{burt1}
\end{equation}
where $\omega$ is the \Kahler{} class. The integer $n_C$ is the
instanton number in the homology class $C$ which we will compute in
the following.

First, note that the prepotential seemingly does not distinguish
torsion homology classes, since $C$ only appears in the integral
$\int_C\omega$ and the integral of a closed form over a torsion
homology class vanishes. However, this neglects the fact that $\omega$
is the complexified \Kahler{} class
\begin{equation}
  \omega 
  = 
  t^a e_a
  =
  B+\iunit J 
  .
\end{equation}
One can always expand the \Kahler{} form in a basis of harmonic
$(1,1)$-forms $\{e_a\}$, but that is not entirely true for the
$B$-field. To define the topological string on a \Kahler{} manifold,
the $B$-field must be locally closed, $dB=0$. However, this only
requires that its characteristic class $H\in H^3(X,\Z)$ vanishes in
$H^3(X,\R)$. In other words, $B$ need not be globally defined and, in
that case, $\int_C B$ makes no sense.

The correct way to define the instanton contribution to the path
integral, $e^{\iunit S(C)}=e^{2\pi\iunit\int_C\omega}$, is as an abstract
map~\cite{Aspinwall:1994uj}
\begin{equation}
  e^{\iunit S}:~ H_2\big(X,\Z\big) \to \Cunits.
\end{equation}
Such a map is determined by the image of the generators. In the case at
hand, we need $3$ generators for $H_2(X,\Z)_\free=\Z^3$ and $2$ more
for $H_2(X,\Z)_\tors=\Z_3\oplus\Z_3$. We denote the image of the free
generators as
\begin{equation}
  e^{\iunit S}(s_0) = p = e^{2\pi\iunit t^1}
  ,~
  q = e^{2\pi\iunit t^2}
  ,~
  r = e^{2\pi\iunit t^3}
  ,
\end{equation}
where $s_0$ is the curve defined in eq.~\eqref{eq:si}. Finally, we
denote the image of the two torsion generators as $b_1$ and $b_2$.
Since three times any torsion curve class is zero, the torsion generators
satisfy
\begin{equation}
  b_1^3 =b_2^3=1.
\end{equation}
The instanton factor $e^{\iunit S}(C)$ of any curve class 
\begin{equation}
  C = (n_1,n_2,n_3,m_1,m_2) 
  ~ \in \Z^3\oplus\Z_3\oplus\Z_3  
  = H_2\big(X,\Z\big)  
\end{equation}
is then a monomial in these generators, that is,
\begin{equation}
  \label{eq:Cmonomial}
  e^{\iunit S}(C) = p^{n_1} q^{n_2} r^{n_3} b_1^{m_1} b_2^{m_2}.
\end{equation} 
It follows that the genus-$0$ non-perturbative superpotential
eq.~\eqref{burt1} is a power-series expansion in $p,q,r,b_{1},b_{2}$
given by
\begin{equation}
  \FprepotX^{np} =  
  \sum_{C\in H_2(X,\Z)} n_C \; 
  \Li_3\Big(p^{n_1} q^{n_2} r^{n_3} b_1^{m_1} b_2^{m_2}\Big).
  \label{burt2}
\end{equation}

\section{Counting Curves}
\label{eq:count}

To explicitly compute the instanton number for each curve class, one
must evaluate the non-perturbative superpotential in the A-model
topological field theory on $X$. After expanding as a power series in
the generators, the instanton numbers can be read off by comparing
with eq.~\eqref{burt2}.  Although the general form of the expansion
for non-simply connected Calabi-Yau threefolds with torsion has been
known for a long time~\cite{Aspinwall:1994uj}, no such example has
ever been explicitly computed. The obvious guess as to how to do this
would be to use mirror symmetry. However, it is not known how to deal
with torsion curves in this context.

To compute the non-perturbative prepotential on $X$, we use a
combination of different techniques. To begin with, we evaluate the
prepotential directly in the A-model as follows.
\begin{itemize}
\item Part of the prepotential on $\Xt$ is known
  analytically~\cite{Hosono:1997hp}. Quotienting this result, one can
  calculate the prepotential on $X$ to all orders in $q, r, b_1,
  b_{2}$, but only to $O(p)$.
\item We also compute the same part of the prepotential by a direct
  A-model computation on $X$.
\end{itemize}
The detailing calculations will be given in~\cite{PartA}. Here, we
simply present the results. We find that
\begin{equation}
  \label{eq:PrepotX}
  \begin{split}
    \FprepotXNP
    (p,q,r,b_1,b_2)
    =&\;
    \left( \sum_{i,j=0}^2 p b_1^i b_2^j \right)
    P(q)^4
    P(r)^4
    +O(p^2)
    \\
   =&\;
  p
  (1+b_1+b_1^2)
   (1+b_2+b_2^2)
    P(q)^4
   P(r)^4
    +O(p^2)
   .
 \end{split}
\end{equation}
Expanding this as a power series and comparing it to
eq.~\eqref{burt2}, one can read off the instanton numbers. The
constant part in $p$ vanishes, so
\begin{equation}
  n_{(0,n_2,n_3,m_1,m_2)} = 0 
  \quad \forall n_2,n_3\in\Z,\, m_1,m_2\in \Z_3
  .
\end{equation}
At linear order in $p$, that is, $n_{1}=1$, the instanton numbers are
non-vanishing. However, they do not depend on the torsion part of the
homology class. That is,
\begin{equation}
  n_{(1,n_2,n_3,m_1,m_2)}
  =
  n_{(1,n_2,n_3,0,0)}
  \quad 
  \forall m_1,m_2\in \{0,1,2\}
  .
\end{equation}
We list the instanton numbers for $n_{2},n_{3} \leq 5$
in~\autoref{tab:1qrb1b2Inst}.
\begin{table}[htpb]
  \centering
  \renewcommand{\arraystretch}{1.3}
  \newcommand{\s}{}
  \begin{tabular}{c|cccccc}
    \backslashbox{$\mathrlap{n_2}$}{$\mathclap{n_3~}$}
    &
    $0$ & $1$ & $2$ & $3$ & $4$ & $5$
    \\ \hline
    $0$ &
    $1$&$4$&$14$&$40$&$105$&$252$
    \\
    $1$ &
    $4$&$16$&$56$&$160$&$420$&$\s1008$
    \\
    $2$ &
    $14$&$56$&$196$&$560$&$\s1470$&$\s3528$
    \\
    $3$ &
    $40$&$160$&$560$&$\s1600$&$\s4200$&$\s10080$
    \\
    $4$ &
    $105$&$420$&$\s1470$&$\s4200$&$\s11025$&$\s26460$
    \\
    $5$ &
    $252$&$\s1008$&$\s3528$&$\s10080$&$\s26460$&$\s63504$
  \end{tabular}
  \caption{Instanton numbers $n_{(1,n_2,n_3,\ast,\ast)}$ 
    computable in the A-model. In this case (for $n_1=1$), 
    the instanton number is independent of the 
    torsion part of the homology class.}
  \label{tab:1qrb1b2Inst}
\end{table}
An important conclusion can be drawn from these results. The
$9$ curves $s_0,\dots,s_8$ of eq.~\eqref{eq:si} contribute at degree
$p b_1^i b_2^j$. We see from~\autoref{tab:1qrb1b2Inst} that
\begin{equation}
  \label{burt3}
  n_{(1,0,0,m_{1},m_{2})}=1
  , \quad  
  \forall m_1,m_2\in \{0,1,2\}
  .
\end{equation}
Hence, each of the curves $s_0,\dots,s_8$ is the lone genus-$0$ curve
in their respective homology class. It follows from this that the
worldsheet instanton induced non-perturbative superpotential can
\emph{not} vanish in heterotic standard model vacua. The explicit
superpotential in these theories will be computed elsewhere.  Note
that had we ignored that torsion subgroup in the degree-$2$ integer
homology, the number of instantons in the $(1,0,0)$ homology class
would have been found to be $n_{(1,0,0)}=9$. One would then be unable
to reach any conclusion about whether or not the superpotential
vanished. Hence, it is essential that the torsion information be
explicitly included when computing the instanton numbers.

Although these results are already significant, one would like to
extend them to any order in the modulus $p$, not just linear order.
This can be accomplished in two ways.
\begin{itemize}
\item While a direct application of mirror symmetry to the covering
  space $\Xt$ fails (there are not enough toric divisors), one can
  compute the prepotential on the Batyrev-Borisov mirror
  $\Xt^\ast$~\cite{Batyrev:1994pg}. This can be used to determine any
  desired term in the A-model prepotential of $X$, but is
  computationally very intensive.
\item The toric mirror of the partial quotient $\Xt/\Z_3$ is
  manageable, and can be used to compute the expansion of the A-model
  prepotential on $X$ to any desired order.
\end{itemize}
Using these techniques, one can obtain the expansion of the
non-perturbative prepotential to any order in each of the generators. The
explicit calculations will be given in~\cite{PartB}. Here, we
simply present some of the results. We find that up to total degree
$4$ in $p,q,r$, the genus-$0$ non-perturbative prepotential is
\begin{equation}
  \begin{split}
    \FprepotX^{np}
    &(p,q,r,b_1,b_2) =
    \\ &\;+
    \sum_{i,j=0}^2
    \Big(
    \begin{array}[t]{ll}
        \Li_3(p b_1^i b_2^j)
      + 4  \Li_3(p q b_1^i b_2^j)
      + 4  \Li_3(p r b_1^i b_2^j)
      \\~
      + 14  \Li_3(p q^2 b_1^i b_2^j)
      + 16  \Li_3(p q r b_1^i b_2^j)
      + 14  \Li_3(p r^2 b_1^i b_2^j)
      \\~
      + 40  \Li_3(p q^3 b_1^i b_2^j)
      + 56  \Li_3(p q^2 r b_1^i b_2^j)
      + 56  \Li_3(p q r^2 b_1^i b_2^j)
      \\~
      + 40  \Li_3(p r^3 b_1^i b_2^j)
      -2 \Li_3(p^2 q b_1^i b_2^j)
      -2 \Li_3(p^2 r b_1^i b_2^j)
      \\~
      -28 \Li_3(p^2 q^2 b_1^i b_2^j)
      +32 \Li_3(p^2 q r b_1^i b_2^j)
      -28 \Li_3(p^2 r^2 b_1^i b_2^j)
      \hspace{2ex}\smash{\Big)}
    \end{array}
    \\ &\;+
    3 \Li_3(p^3 q ) + 3 \Li_3(p^3 r)
    \\ &\;+ 
    (\text{degree }\geq 5)
    .
  \end{split}  
\end{equation} 
The instanton numbers can be read off by comparing this expansion to
eq.~\eqref{burt2}. First, we find that the order $p$ instanton numbers
obtained from this expansion are identical to those given
in~\autoref{tab:1qrb1b2Inst}, as they must be.  Second, the order
$p^{2}$ instanton numbers do not depend on the torsion part of the
homology class, as was the case at order $p$.

Some illustrative results at order $p^{3}$ are given
in~\autoref{tab:n1n2n3b1b2Inst}.
\begin{table}[htpb]
  \centering
  \renewcommand{\arraystretch}{1.3}
  \newcommand{\s}{\scriptstyle}
  \newcommand{\sss}{\hspace{5mm}}
  \begin{tabular}{@{\sss}c@{\sss}@{\hspace{10mm}}@{\sss}c@{\sss}}
    $n_{(3,n_2,n_3,0,0)}$ &
    $n_{(3,n_2,n_3,m_1,m_2)},~(m_1,m_2)\not=(0,0)$
    \\
    \begin{tabular}{c|ccc}
      \backslashbox{$\mathrlap{n_2}$}{$\mathclap{n_3~}$}
      &
      $0$ & $1$ & $2$ 
      \\ \hline
      $0$ &
      $0$&$\mathemph{3}$&$\mathemph{36}$
      \\
      $1$ &
      $\mathemph{3}$&$\mathemph{108}$
      \\
      $2$ &
      $\mathemph{36}$
    \end{tabular}
    &
    \begin{tabular}{c|ccc}
      \backslashbox{$\mathrlap{n_2}$}{$\mathclap{n_3~}$}
      &
      $0$ & $1$ & $2$ 
      \\ \hline
      $0$ &
      $0$&$\mathemph{0}$&$\mathemph{27}$
      \\
      $1$ &
      $\mathemph{0}$&$\mathemph{81}$
      \\
      $2$ &
      $\mathemph{27}$
    \end{tabular}
  \end{tabular}
  \caption{Some of the instanton numbers $n_{(n_1,n_2,n_3,m_1,m_2)}$ computed 
    by mirror symmetry. The entries marked in
    \textcolor{red}{\textbf{bold}} depend non-trivially 
    on the torsion part of their 
    respective homology class.}
  \label{tab:n1n2n3b1b2Inst}
\end{table}
Interestingly, we observe that at order $p^3$ the number of instantons
\emph{does} depend on the torsion part of their homology class. Note,
for example, that there are $3$ instantons in the homology class
$(3,1,0,0,0)\in H_2(X,\Z)$, but there are no instantons in the other
$8$ curve classes $(3,1,0,i,j)\in H_2(X,\Z)$ that differ in their
torsion part.

\section*{Acknowledgments}

The authors would like to thank Albrecht Klemm, Tony Pantev, and
Masa-Hiko Saito for valuable discussions. 
\GrantsAcknowledgements
V.~B. would like to thank the Texas A\&M University for the
opportunity to talk about this work.

\bibliographystyle{utcaps} \renewcommand{\refname}{Bibliography}
\addcontentsline{toc}{section}{Bibliography} 
\small
\bibliography{Volker,Emanuel}

\end{document}